\documentclass[twocolumn,
superscriptaddress,amsmath, amssymb, amsfonts,preprintnumbers,aps,prd,
longbibliography,nofootinbib]{revtex4-2}

\usepackage{scalerel}
\usepackage{color}
\usepackage{amsmath,amsfonts,amssymb}%,calrsfs}
\usepackage{slashed}
\usepackage{latexsym,epsfig}
\usepackage{dsfont}
\usepackage{arydshln}
\usepackage{extarrows}
\usepackage{hyperref}
\usepackage{array}
\def\moth{\mathsurround=0pt}
\newdimen\zo \zo=0pt

\def\tick{\leaders\hrule height 0.5ex depth 0pt \hskip 0.5pt}
\def\upboxfill{$\moth \setbox\zo\hbox{\tick}%
  \hskip 3pt\hbox to 0pt{$\tick$\hss}\hrulefill \hbox to 7.5pt{$\tick$\hss}$}

\def\dtick{\leaders\hrule height .34pt depth 0.5ex \hskip 0.5pt}
\def\downboxfill{$\moth \setbox\zo\hbox{\dtick}%
  \hskip 2pt\hbox to 0pt{$\dtick$\hss}\hrulefill \hbox to 2pt{$\dtick$\hss}$}

\def\ov{\bar}

\def\p{\partial}

\def\bec{\begin{center}}
\def\ec{\end{center}}

\def\ov{\overline}

\def\nn{\nonumber}

\def\be{\begin{equation}}
\def\ee{\end{equation}}
\def\bea{\begin{eqnarray}}
\def\eea{\end{eqnarray}}
\def\ba{\begin{array}}
\def\ea{\end{array}}

\begin{document}

\title{Non-commutative double geometry}

\author{Toni Kodžoman} 
\email{tkodzoman@irb.hr}
\affiliation{Division of Theoretical Physics, Rudjer Boskovic Institute, Bijenicka 54, 10000 Zagreb, Croatia}

\author{Eric Lescano}
\email{elescano@irb.hr}
\affiliation{University of Wroclaw, Faculty of Physics and Astronomy, Maksa Borna 9, 50-204 Wroclaw,
Poland}

\affiliation{Division of Theoretical Physics, Rudjer Boskovic Institute, Bijenicka 54, 10000 Zagreb, Croatia}

\begin{abstract} 

\noindent We construct non-commutative theories with the Moyal-Weyl product in the Double Field Theory (DFT) framework. We deform the infinitesimal generalized diffeomorphisms and the Leibniz rule in a consistent way. The prescription requires a generalized star metric, which can be thought of as the fundamental double metric, in order to construct the action. Finally we use the generalized scalar field dynamics and the generalized scalar field-perfect fluid correspondence to construct the generalized energy momentum-tensor of a perfect fluid in the non-commutative double geometry. The present formalism paves the way to the study of string cosmologies scenarios including the Moyal-Weyl product in a T-duality invariant way.    

\end{abstract}

\maketitle

\section{Introduction}

T-duality is an exact symmetry of closed string theory \cite{Ashoke} which establishes that toroidal backgrounds $T^d$ related via the non-compact group $O(d,d,Z)$ are physically equivalent. When one compactifies the low energy limit of string theory (ST) considering Kaluza-Klein compactifications, the continuous form of the duality group, $O(d,d,R)$, appears as a symmetry. Interestingly enough, the effect of T-duality is producing a non-commutative \cite{nc1}-\cite{strings9} relation in the commutator of the target coordinates and/or its dual coordinates \cite{intrinsic1}-\cite{intrinsic2}. This effect can be thought of as a central extension of the zero-mode operator algebra, an effect set by the string length scale even in cases where the backgrounds are trivial.

\noindent At the effective level (supergravity level) it is possible to construct commutative theories where the T-duality group appears as a symmetry before compactification by doubling the geometry and (re-)writing all the fundamental fields and parameters via multiplets of the duality group. This framework is known as Double Field Theory (DFT) \cite{Siegel1}-\cite{DFT6}. One interesting aspect about DFT is the existence of a strong constraint or section condition which can be performed in order to get rid of the extra coordinates required by $O(D,D,R)$ invariance, where $D=n+d$, and $n$ is the dimension of the external space. Applying the strong constraint only on the external coordinates it is possible to arrive to a hybrid theory with an $n$-dimensional exterior space-time and a $d$-dimensional double internal space \cite{HandS}. While this is a promising formulation constructed on an arbitrary double internal space, the formulation is purely classical. While this setup can be used to effectively describe theories and its duals, the relation between non-commutativity and gravity has been studied in the double geometry. Another missing ingredient in this type of formulation is the possibility that both the external and internal backgrounds are partially/fully non-Riemannian \cite{ParkNR1}-\cite{NR5}. 

\noindent  If we focus on a model which could be reduced to the Einstein-Hilbert
action, one possibility is to define a non-commutative geometry considering that the commutator of the coordinates is given by \cite{book}
\bea
\Big[x^{\mu},x^{\nu} \Big] = i \theta^{\mu \nu} \, ,
\eea
with $\theta$ a generic antisymmetric tensor which is also used to defined an algebra ${\cal A}_{\theta}$ through the following star product,
\begin{equation}
f\star g\,(x) = e^{\frac{i}{2}\theta^{\rho\sigma} \frac{\partial }{\partial x^\rho} \otimes
\frac{\partial }{\partial y^\sigma}}f(x) \otimes g(y)
\Big|_{y\rightarrow x} \, ,
\end{equation}
$\mu= 1,\dots, D-1$. In this work we will focus on the $\theta=\textit{const.}$ case (Moyal-Weyl product), where Lorentz invariance is broken from the very beginning \cite{Harikumar}. Since the ordinary relation 
\bea
g_{\mu \nu} \star g^{\nu \rho} = \delta_{\mu}^{\rho} + \mathcal{O}(\theta) \, ,
\eea
receives $\theta-$corrections, one is forced to construct an inverse metric compatible with the star product, $g^{\star \mu \nu}$, which satisfies
\bea
g_{\mu \nu} \star g^{\star \nu \rho} = \delta_{\mu}^{\rho}  \, .
\eea

\noindent While the natural expectation is to deform both symmetries and action using the star product, the natural proposal
\bea
\delta_{\xi} v^{\mu} = \xi^{\nu} \star \partial_{\nu} v^{\mu} - \partial_{\nu} \xi^{\mu} \star v^{\nu}
\eea
does not satisfy closure
\bea
[\delta_{\xi_1}, \delta_{\xi_2}] = \delta_{\xi_{21}} \, .
\eea

\noindent In \cite{Marija}, the authors keep diffeomorphism transformations untouched (with a modification to the Leibniz rule) and they define a consistent action as
\bea
S = \int d^{D}x \sqrt{- \textrm{\textbf{g}}} \star R \, + \textrm{cc} = \int d^{D}x \sqrt{-g} \, R + \mathcal{O}(\theta^2) ,
\label{action0}
\eea
where both the determinant of the metric and the Ricci scalar contain $\theta-$contributions \cite{book} and the cc (complex conjugate) terms guarantee that the action is real. Therefore the previous action is invariant under infinitesimal diffeomorphisms, paying the cost of deforming the Leibniz rule \cite{Marija}. In this work we deform the transformation rules using a generic commutative and associative product, so the Leibniz rule is consistently modified but the action remains unchanged. Moreover, the closure of the transformations is also fulfilled. For clarity's sake, we present our main results in the following part.    

\subsection{Main results}

Our formulation consists of a systematic procedure to construct $\theta$-deformed theories and we mainly focus on the DFT framework. In this work we provide a simple way to generalize this construction defining the notion of DFT \footnote{See \cite{reviewDFT1}-\cite{reviewDFT3} for pedagogical reviews and \cite{others1}-\cite{others5} for related works.} on non-commutative double spaces. 

\noindent We start by considering the non-commutative relation at the level of double coordinates,
\be
[{X}^M , {X}^N] = i \theta^{M N} \, ,
\ee
where $M=0,\dots,2D-1$. We define the generalized star metric through 
\bea
{\cal H}_{M P}^{\star} \star {\cal H}^{\star P N} = \delta_{M}^{N} \, ,
\eea
since there is a notion of an inverse metric in DFT. The generalized star metric can be expanded in powers of $\theta$ as 
\bea
{\cal H}^{\star P N} = {\cal H}^{P N} - \frac{i}{4} {\cal H}^{P R} \theta^{Q S} \partial_{Q}{\cal H}_{R T} \partial_{S}{\cal H}^{T N} + \mathcal{O}(\theta^2) \, . 
\label{starH}
\eea
Moreover, we consider this new metric as the fundamental field of the theory, so the construction of the non-commutative version of the DFT projectors is given by
\bea
P^{\star}_{MN} = \frac{1}{2}\left(\eta_{MN} - {\cal H}^{\star}_{MN}\right)  \ \ {\rm and} \ \
\ov{P}^{\star}_{MN} = \frac{1}{2}\left(\eta_{MN} + {\cal H}^{\star}_{MN}\right)\ ,
\eea
which satisfy the following properties 
\bea
&& {\overline{P}}^{\star}_{{M Q}} \star {\overline{P}}^{\star Q}{}_{ N}={\overline{P}}^{\star}_{{M N}}\, , \\ && {P}^{\star}_{{M Q}} \star {P}^{\star Q}{}_{ N}={P}^{\star}_{{M N}}, \nn\\
&& {P}^{\star}_{{M  Q}} \star {\overline{P}}^{\star Q}{}_{ N} = {\overline{P}}^{\star}_{ {M Q}} \star  {P}^{\star Q}{}_{ N} = 0 \,.
\eea
Mimicking the procedure given in \cite{OZ}, we define a connection $\Gamma_{M N P}$ and a covariant Riemann tensor as
\bea
{\cal R}_{MNKL} = R_{MNKL} + R_{KLMN} + \Gamma_{QMN} \star \Gamma^{Q}{}_{KL} \, , \\
{R}_{MNKL} = 2 \partial_{[M} \Gamma_{N]K L} + 2 \Gamma_{[M| Q L} \star \Gamma_{N]K}{}^{Q} \, .
\eea 

\noindent The action is given by
\bea
\int d^{2D}X \, e^{-2\textrm{\textbf{d}}} \star  P^{\star M N} \star P^{\star Q R} \star {\cal R}_{M Q N R} + \textrm{cc} \, .
\eea

\noindent Finally, we have checked that the inclusion of matter through a generalized and massless scalar field is also consistent with our formalism
\bea
{\cal L}_{\textrm{matter}}[ {\cal H}, \Phi] = \frac12   \partial_{ M}  \Phi \star {\cal H}^{\star M  N} \star \partial_{ N}  \Phi \, - V(\Phi) ,
\eea
which is compatible with the free scalar action to all orders in $\theta$ after parametrization and imposing the strong constraint. We consider the generalized perfect fluid-scalar field correspondence in order to construct the generalized energy momentum tensor 
\bea
{\cal T}_{ M  N} & = & 4\,\Big[\overline{P}^{\star}_{[M|K} \star P^{\star}_{N]L}\Big](\sqrt{\tilde e + \tilde p} \ U_{M} \star \sqrt{\tilde e + \tilde p} \ U_{N}) \nn \\ && - \frac{1}{2} \eta_{M N} \sqrt{\tilde e + \tilde p} \ U_{P} \star {\cal H}^{\star P Q} \star \sqrt{\tilde e + \tilde p} \ U_{Q} \, ,
\label{emintro}
\eea
with $U_{M}$ a generalized velocity \cite{EyN2} and $\tilde e$, $\tilde p$ the generalized energy density and pressure respectively. The previous tensor can be used in order to construct the generalized Einstein equation for these geometries, which pave the way to the study of non-commutativity string cosmologies in a T-duality invariant way. 

\section{Non-commutative geometry with Moyal-Weyl product}
\label{sectionNC}

\subsection{Symmetry transformations}

The algebra
${\cal{A}}_\theta$ is based on the relation
\be
[{x}^\mu , {x}^\nu] = i \theta^{\mu\nu}
\ee
with $\theta^{\mu\nu}$ constant and real. This algebra can be realized on the linear space ${\cal{F}}$ of complex functions $f(x)$ of commuting variables: The elements of the algebra ${\cal{A}}_\theta$ are represented by functions of the commuting variables
$f(x)$, their product by the Moyal-Weyl star product, i.e.,
\begin{equation}
f\star g\,(x) = e^{\frac{i}{2} \theta^{\rho\sigma} \frac{\partial }{\partial x^\rho} \otimes
\frac{\partial }{\partial y^\sigma}}f(x) \otimes g(y)
\Big|_{y\rightarrow x} \, .
\end{equation}

\noindent The $\star$-derivative, $\p^\star_{\mu} f := \p_\mu f$ , satisfies 
\begin{equation}
\p_{\mu}  x^\rho =\delta _\mu^\rho ,
\end{equation}
and the usual product rule with respect to the $\star$-product (from now on we use the standard notation for the star derivative),
\begin{equation}
\p_\mu (f\star g) = (\p_\mu f)\star g +f\star (\p_\mu g) .
\end{equation}

\noindent Finally let us construct the Leibniz rule. It is possible to define the infinitesimal diffeomorphisms in the following form for scalars fields
\begin{equation}
\hat{\delta}_\xi \phi = X^\star_\xi\triangleright \phi ,
\end{equation}
and for vector fields
\begin{equation}
\hat{\delta}_\xi V^\mu = X^\star_\xi \triangleright V^\mu + X^\star_{(\p_\rho \xi^\mu)}\triangleright V^\rho .
\end{equation}
The operators $X^\star_\xi$ and $X^\star_{(\p_\mu \xi^\lambda)}$ are given by
\footnotesize
\begin{eqnarray}
X^\star_\xi \hspace*{-2mm}&=&\hspace*{-2mm}\sum_{n=0}^\infty \frac{1}{n!}\Big(-\frac{i}{2}\Big)^n
\theta^{\rho_1\sigma_1}\dots\theta^{\rho_n\sigma_n}(\partial_{\rho_1}\dots\partial_{\rho_n}\xi^\lambda)
\star \partial^\star_{\sigma_1}\dots\partial^\star_{\sigma_n}\p^\star_\lambda , \nn \\
X^\star_{(\p_\mu \xi^\lambda)} \hspace*{-2mm}&=&\hspace*{-2mm}
\sum_{n=0}^\infty \frac{1}{n!}\Big(-\frac{i}{2}\Big)^n
\theta^{\rho_1\sigma_1}\dots\theta^{\rho_n\sigma_n}(\partial_{\rho_1}\dots\partial_{\rho_n}\p_\mu\xi^\lambda)
\star \partial^\star_{\sigma_1}\dots\partial^\star_{\sigma_n}  \, .  \nn
\end{eqnarray}
\normalsize
The deformed Leibniz rule takes the following form, 
\bea
X^\star_\xi \triangleright (f\star g) & = & \mu_\star \{ e^{-\frac{i}{2}\theta^{\rho\sigma}\p^{\star}_{\rho} \otimes\p^{\star}_{\sigma}}
\Big( X^\star_\xi\otimes 1 \nn \\ && + 1\otimes X^\star_\xi\Big)
e^{\frac{i}{2}\theta^{\rho\sigma}\p^{\star}_{\rho} \otimes \p^{\star}_{\sigma}} \triangleright (f\otimes g) \}. 
\eea

\noindent So far we have reviewed the basis of non-commutativity in flat space. It proves illustrative to focus first on the physics of a free massless scalar field in non-commutative geometry. The action in this case is given by
\bea
S_{\phi} & = & \frac12 \int d^{D}x  \partial_{\mu} \phi \star \partial^{\mu} \phi \label{scalargr}
\eea
where the indices are contracted with a flat inverse metric $\eta^{\mu \nu}$.

\noindent From the previous Lagrangian, it is simple to observe that the terms created for the $\theta$-expansion are even and real. Moreover, in this case the squared contribution is a total derivative due to the constant and antisymmetric nature of $\theta$. The most interesting point here is that the $\phi$-field is no longer a scalar field when the star transformations are expanded. Therefore, we can analyze covariance with respect to the star product or with respect to the ordinary product. While $\phi$ is a scalar for the former, for the latter its transformation is non-covariant. However, it is straightforward to prove that the action (\ref{scalargr}) is invariant. In these geometries, Lorentz invariance is broken from the very beginning. For instance, taking a 4-dimensional manifold, if the only non-vanishing component of $\theta^{\mu \nu}$ is the $\theta^{12}$-component then Lorentz boosts in the $3$-direction and rotations in the $1$-$2$ plane are still preserved as are translations in any
direction. 

\noindent In the next section we will review the basic steps to generalize these concepts to the gravitational case.

\subsection{Non-commutative gravity}

For more general scenarios we need the notion of a covariant derivative. The introduction of this operator follows as in the commutative case, in the sense that we introduce a connection $\Gamma_{\mu\nu}^{\rho}$ in order to define the covariant derivative as
\begin{equation}
\nabla_{\mu}V_{\nu} := \partial_{\mu} V_{\nu}-\Gamma_{\mu\nu}^{\rho}\star V_{\rho}. 
\end{equation}

\noindent The metric of a non-commutative $g_{\mu \nu}$ is defined as a symmetric tensor under infinitesimal diffeomorphisms, and its inverse $g^{\star \mu \nu}$ is constructed such that 
\bea
g_{\mu \nu} \star g^{\star \nu \rho} = \delta_{\mu}^{\rho} \, .
\eea

\noindent Therefore $g^{\star \mu \nu}$ is not symmetric and the non-covariant contributions compensate the non-covariant terms arising from the star product, 
\bea
g^{\star \mu \nu} = g^{\mu \nu} - \frac{i}{2} g^{\mu \xi} \theta^{\rho \sigma} \partial_{\rho} g_{\xi \epsilon} \partial_{\sigma} g^{\epsilon \nu} + \mathcal{O}(\theta^2) \, .
\eea

\noindent Using the previous metrics it is possible to determine the torsionless star Levi-Civita connection, defined as,
\bea
\Gamma_{\nu \beta}^{\sigma} =  \frac12 \Gamma_{\gamma \nu \beta} \star g^{\star \gamma \sigma} =  \frac12(\partial_{\nu} g_{\beta \gamma} + \partial_{\beta} g_{\nu \gamma} - \partial_{\gamma} g_{\nu \beta}) \star g^{\star \gamma \sigma}
\eea
while the Riemann tensor is given by
\bea
R_{\mu \nu \sigma}{}^{\rho} = 2 \partial_{[\mu} \Gamma_{\nu] \sigma}^{\rho} + 2 \Gamma_{\kappa [\mu}^{\rho} \star \Gamma_{\nu] \sigma}^{\kappa} \, .
\eea
From the previous expression we can extract the Ricci scalar for the non-commutative geometries as
\bea
R & = & g^{\star \mu \nu} \star R_{\nu \sigma \mu}{}^{\sigma} \, ,
\eea
which transforms as a scalar. Finally, the action can be defined as
\bea
S=\int d^Dx \sqrt{-\textrm{\textbf g}} \star R + \textrm{cc} \, ,
\eea
where the determinant of the metric also includes star contributions \cite{Marija}. 

\section{Double Field Theory: the commutative case}
\label{sectionDFT}

The geometry of DFT is based on a double space equipped with $\eta_{M N}$, an invariant metric of $O(D,D)$ and a dynamical metric,  the generalized metric ${\cal H}_{M N}$. The indices $M,N,\dots$ are in the fundamental representation of the duality group and are raised and lowered with $\eta^{M N}$ and $\eta_{M N}$ respectively. The generalized metric ${\cal H}_{M N}$ encodes the bosonic tensors of the universal NS-NS sector of string theory, namely, a metric tensor $g_{\mu \nu}$ and a b-field $b_{\mu \nu}$. This metric is a tensor under $O(D,D)$ transformations. Infinitesimal $O(D,D)$-transformations acting on an arbitrary double vector read
\bea
\delta_h V^M = V^{N} h_{N}{}^{M} \, ,
\label{duality}
\eea
where $h \in O(D, D)$ is an arbitrary parameter. Another symmetry of the theory is given by generalized diffeomorphisms, generated infinitesimally  by $\xi^{M}$  through the generalized Lie derivative, defined by 
\bea
{\cal L}_\xi V_M = \xi^{N} \partial_N V_M + (\partial_M \xi^N - \partial^N \xi_{M}) V_N  \, ,
\eea 
where $V_M$ is an arbitrary generalized vector. The closure of the gauge transformations 
\bea
\Big[\delta_{\xi_1},\delta_{\xi_2} \Big] = \delta_{\xi_{21}} \,,
\eea
is given by the C-bracket
\bea
\xi^{M}_{12}(X) = \xi^{P}_{1} \frac{\partial \xi^{M}_{2}}{\partial X^{P}} - \frac12 \xi^{P}_{1} \frac{\partial \xi_{2P}}{\partial X_{M}} - (1 \leftrightarrow 2) \, .
\eea
 The DFT Jacobiator is not trivial (but it is given by a trivial parameter) and therefore the algebraic structure of DFT is given by an $L_{\infty}$-algebra with a non-trivial $l_3$ product \cite{Linf1}-\cite{Linf4}, which measures the failure of the Jacobi identity in the double geometry.

\noindent On the other hand, the generalized metric is an element of $O(D,D)$ and therefore satisfies
\bea
{\cal H}_{M P} {\cal H}^{P N} = \delta_{M}^{N} \, .
\label{Odd}
\eea
Using both DFT metrics one can construct the DFT projectors in the following way
\bea
P_{MN} = \frac{1}{2}\left(\eta_{MN} - {\cal  H}_{MN}\right) \, , \  
\ov{P}_{MN} = \frac{1}{2}\left(\eta_{MN} + {\cal H}_{MN}\right)\ .
\label{proyc}
\eea
The previous projectors satisfy
\bea
{\overline{P}}_{{M Q}} {\overline{P}}^{ Q}{}_{ N}& = &{\overline{P}}_{{M N}}\, , \quad {P}_{{M Q}} {P}^{Q}{}_{ N}={P}_{{M N}}, \nn\\
{P}_{{M  Q}}{\overline{P}}^{Q}{}_{ N} & = & {\overline{P}}_{ {M Q}}  {P}^{ Q}{}_{ N} = 0\, .  \
\label{curveprojrel}
\eea
One of the purposes of DFT is to define a theory manifestly invariant under $O(D,D)$, which is a symmetry of string theory. Because of that, all the DFT fields and parameters are $O(D,D)$ multiplets or group-invariant objects. Since the dimension of the fundamental representation of $O(D,D)$ is $2D$, the coordinates of DFT are $X^{M}=(x^{\mu},\tilde{x}_{\mu})$. The coordinates $\tilde{x}_{\mu}$ are known as the dual coordinates and are taken away imposing the strong constraint,
\be
\partial_{M} (\partial^{M} V) = (\partial_{M} V) (\partial^{M} W) = 0 \, ,
\label{SC}
\ee
where $V$ and $W$ can be products of arbitrary generalized fields or parameters. Solving the previous constraint with $\tilde \partial^{\mu}=0$, the components of the fields of DFT depend only on $x^{\mu}$. The parametrization of the invariant metric is given by, 
\bea
{\eta}_{{M N}}  = \left(\begin{matrix}0&\delta_\nu^\mu\\ 
\delta^\nu_\mu&0 \end{matrix}\right)\, ,  \label{eta}
\eea
while the parametrization of the generalized metric is given by
\bea
{\cal H}_{{M N}}  = \left(\begin{matrix}g^{\mu \nu}&-g^{\mu \sigma}b_{\sigma \nu}\\ 
b_{\mu \sigma}g^{\sigma \nu}&g_{\mu \nu}-b_{\mu \sigma} g^{\sigma \rho} b_{\rho \nu} \end{matrix}\right)\, , 
\label{Gmetric}
\eea
where $b_{\mu \nu}$ is the Kalb-Ramond field. It is straightforward to check that the previous parametrization satisfies (\ref{Odd}).

\noindent The action of DFT is constructed from the following Lagrangian,
\bea
{\cal L} = && \frac18 {\cal H}^{MN} \partial_{M}{\cal H}^{KL} \partial_{N}{\cal H}_{KL} - \frac12 {\cal H}^{MN} \partial_{N}{\cal H}^{KL} \partial_{L}{\cal H}_{MK} \nn \\ && + 4 {\cal H}^{MN} \partial_{M}\partial_{N}d + 4 \partial_{M}{\cal H}^{MN} \partial_{N}d \nn \\ && - 4 {\cal H}^{MN} \partial_{M}d \partial_{N}d - \partial_{M} \partial_{N} {\cal H}^{MN} \, ,
\label{LagrangianDFT}
\eea
where $d$ is known as the generalized dilaton and it is parametrized as $\partial_{\nu}{d}=\partial_{\nu} \phi -\frac14 g^{\sigma \rho} \partial_{\nu} g_{\sigma \rho}$. The full action is given by $S=\int e^{-2d} \ {\cal L} \ d^{2D}X$ and after parametrization and using the strong constraint, the resulting action coincides with the low energy limit of the universal NS-NS sector of string theory up to total derivatives,
\bea
S= \int d^{d}X e^{-2\phi} \sqrt{g} (R + 4 \partial_{\mu}\phi \partial^{\mu} \phi - \frac{1}{12} H_{\mu \nu \rho} H^{\mu \nu \rho}) \, ,
\eea
where $H_{\mu \nu \rho}=3\partial_{[\mu} b_{\nu \rho]}$ is the curvature of the Kalb-Ramond field.

\section{Non-commutative Double Field Theory}
\label{sectionNCDFT}

\subsection{Vacuum fields and action}

The fundamental non-commutative relation between double coordinates is defined as
\be
[{X}^M , {X}^N] = i \theta^{M N}
\ee
with $\theta^{M N}$ an $O(D,D)$ real multiplet. This algebra can be realized on the linear space ${\cal{F}}$ of complex functions $F(x)$ of commuting variables: The elements of the algebra ${\cal{A}}_\theta$ are represented by functions of the commuting variables
$F(X)$, their product by the Moyal-Weyl star product, i.e.,
\begin{equation}
F\star G\,(X) = e^{\frac{i}{2} \theta^{P Q} \frac{\partial }{\partial X^P} \otimes
\frac{\partial }{\partial y^Q}}F(X) \otimes G(Y)
\Big|_{Y\rightarrow X} \, ,
\end{equation}
where $\theta^{PQ}$ is constant.

\noindent The previous product reproduces the ordinary star product if 
\bea
\frac{\partial }{\partial X^P}
\theta^{P Q}\frac{\partial }{\partial y^Q}
\Big|_{Y\rightarrow X} \rightarrow \frac{\partial }{\partial x^\rho}
\theta^{\rho \sigma}\frac{\partial }{\partial y^\sigma}\Big|_{y\rightarrow x} \, ,
\eea
and therefore the relevant component of $\theta^{M N}$ is given by $M={}^\mu$ and $N={}^\nu$ due to the strong constraint. 

\noindent On the other hand, the $\star$-derivatives satisfies 
\begin{equation}
\p _M  X^P =\delta _M^P ,
\end{equation}
and the usual product rule with respect to the $\star$-product,
\begin{equation}
\p_M (F\star G) = (\p_M F)\star G +F\star (\p_M G) .
\end{equation}

\noindent We impose a deformed Leibniz rule considering that the generalized transformations can be defined as
\begin{equation}
\hat{\delta}_\xi \Phi = X^\star_\xi\triangleright \Phi ,
\end{equation}
for a generalizied scalar field and
\begin{equation}
\hat{\delta}_\xi V^M = X^\star_\xi \triangleright V^M + X^\star_{(\p_P \xi^M)}\triangleright V^P - X^\star_{(\p^P \xi_M)}\triangleright V^P ,
\label{rewriting}
\end{equation}
for a generalized vector field. 

\noindent The generalized metric for non-commutative DFT is given by
\bea
{\cal H}_{{M N}}  = \left(\begin{matrix}g^{\mu \nu}&-g^{\mu \sigma} b_{\sigma \nu}\\ 
b_{\mu \sigma} g^{\sigma \nu}&g_{\mu \nu}-b_{\mu \sigma} g^{\sigma \rho} b_{\rho \nu} \end{matrix}\right)\, , 
\label{Hmetric}
\eea

\noindent and we define the generalized star metric through 
\bea
{\cal H}_{M P}^{\star} \star {\cal H}^{\star P N} = \delta_{M}^{N} \, ,
\label{Oddnc}
\eea
which reduces to the condition (\ref{Odd}) for $\theta=0$. The ${\cal H}^{\star P N}$ metric now contains its own non-covariant $\theta$-expansion as
\bea
{\cal H}^{\star P N} = {\cal H}^{P N} - \frac{i}{4} {\cal H}^{P R} \theta^{Q S} \partial_{Q}{\cal H}_{R T} \partial_{S}{\cal H}^{T N} + \mathcal{O}(\theta^2) \, . 
\label{stargen}
\eea

\noindent The covariant derivative acting on a generic double vector is defined as
\bea
\nabla_{M} V_{N} = \partial_{M} V_{N} - \Gamma_{M N}{}^{P} \star V_{P} \, ,
\label{covderdft}
\eea
where we have introduced a generalized affine connection $\Gamma_{M N}{}^{P}$ whose transformation properties must compensate the failure of the partial derivative of a tensor to transform covariantly under
generalized diffeomorphisms. 

\noindent We can now demand some properties on the connection, namely:
\begin{itemize}
\item \underline{compatibility with $\eta_{MN}$}:
\bea
\nabla_{M} \eta_{N P} = 0 \, ,
\eea
and then the generalized affine connection is antisymmetric in its last two indices, {\it i.e.}
\bea
\Gamma_{M N P} = - \Gamma_{M P N} \, .
\eea
\item \underline{compatibility with ${\cal H}^{\star}_{MN}$}: 
\bea
\nabla_{M} {\cal H}^{\star}_{N P} = 0 \, .
\eea
In order to discuss this item, it is convenient to define the star projectors,
\be
\begin{split}
& P^{\star}_{MN}  = \frac{1}{2}\left(\eta_{MN} - {\cal H}^{\star}_{MN}\right)  \\ 
& \ov{P}^{\star}_{MN} = \frac{1}{2}\left(\eta_{MN} + {\cal H}^{\star}_{MN}\right)\ ,    
\end{split}
\ee
which satisfy the following properties 
\be
\begin{split}
& {\overline{P}}^{\star}_{{M Q}} \star {\overline{P}}^{\star Q}{}_{ N}={\overline{P}}^{\star}_{{M N}}\, , \\  
& {P}^{\star}_{{M Q}} \star {P}^{\star Q}{}_{ N}={P}^{\star}_{{M N}},
\end{split}
\ee 
\be
\begin{split}
& {P}^{\star}_{{M  Q}} \star {\overline{P}}^{\star Q}{}_{ N} = {\overline{P}}^{\star}_{ {M Q}} \star  {P}^{\star Q}{}_{ N} = 0\, ,  \\ 
& {\overline{P}}^{\star}_{{MN}} + {P}^{\star}_{{M N}} = \eta_{{M N}}\,. 
\end{split}
\ee
The projections $\Gamma_{\overline{MNP}}$ and $\Gamma_{\underline{MNP}}$ remain undetermined after imposing $\nabla_{M}P^{\star}_{NP}=0$ and $\nabla_{M} \ov{P}^{\star}_{NP}=0$ as in the non-commutative case.

\item \underline{vanishing torsion}:
\bea
\Gamma_{[MNP]} = \frac13 \,T_{MNP} = 0 \, .
\eea
Let us observe that the generalized torsion $T_{MNP}$ is antisymmetric in all its indices and transforms as a tensor (unlike $\Gamma_{[MN]P}$).   
\end{itemize}

\noindent The non-commutative version of the DFT Lagrangian is given by the following scalar,
\bea
{\cal R} = P^{\star M N} P^{\star Q R} \star {\cal R}_{M Q N R} \, ,
\eea 
 where 
\bea
{\cal R}_{MNKL} = R_{MNKL} + R_{KLMN} + \Gamma_{QMN} \star \Gamma^{Q}{}_{KL} \\
{R}_{MNKL} = 2 \partial_{[M} \Gamma_{N]K L} + 2 \Gamma_{[M| Q L} \star \Gamma_{N]K}{}^{Q}
\eea 
 while the full action can be written as
\bea
S = \int d^{2D}X \, e^{-2d} \star {\cal R} + \textrm{cc}\, ,
\eea
where $e^{-2d} = e^{-2 \phi} \sqrt{\textrm{\textbf{g}}}$. On the other hand the components of the generalized star metric can be easily computed order by order. For example, the first order contributions are given by
\begin{widetext}
\bea
{\cal H}^{\star \rho \nu} & = & g^{\rho \nu} - \frac{i}{4} \theta^{\mu \sigma} \partial_{\mu}g_{\tau \lambda} \partial_{\sigma} g^{\nu \tau} g^{\rho \lambda} - \frac{i}{4} \theta^{\mu \sigma} \partial_{\mu}b_{\tau \lambda} \partial_{\sigma}b_{\xi \delta} g^{\nu \tau} g^{\rho \xi} g^{\lambda \delta} + \mathcal{O}(\theta^2)
\eea
\bea
{\cal H}^{\star \rho}{}_{\nu} & = & - g^{\rho \alpha} b_{\alpha \nu} - \frac{i}{4} \theta^{\mu \sigma} \partial_{\mu} g_{\tau \lambda} \partial_{\sigma}g^{\tau \xi} b_{\nu \xi} g^{\rho \lambda} - \frac{i}{4} \theta^{\mu \sigma} \partial_{\mu}b_{\nu \tau} \partial_{\sigma}g^{\tau \rho} \nn \\ && - \frac{i}{4} \theta^{\mu \sigma} \partial_{\mu} b_{\tau \lambda} \partial_{\sigma}g_{\nu \xi} g^{\rho \tau} g^{\lambda \xi} - \frac{i}{4} \theta^{\mu \sigma} \partial_{\mu} b_{\tau \lambda} \partial_{\sigma}b_{\xi \delta} b_{\nu \alpha} g^{\rho \tau} g^{\lambda \xi} g^{\delta \alpha} + \mathcal{O}(\theta^2) \eea
\bea
{\cal H}^{\star}_{\rho \nu} & = & g_{\rho \nu} - b_{\rho \alpha} g^{\alpha \beta} b_{\beta \nu} - \frac{i}{4} \theta^{\mu \sigma} \partial_{\mu} g_{\tau \lambda} \partial_{\sigma}g^{\tau \xi} b_{\nu \xi} b_{\rho \chi} g^{\lambda \chi} + \frac{i}{4} \theta^{\mu \sigma} \partial_{\mu} b_{\nu \tau} \partial_{\sigma}g_{\lambda \xi} b_{\rho \chi} g^{\tau \lambda} g^{\xi \chi} \nn \\ && + \frac{i}{4} \theta^{\mu \sigma} \partial_{\mu}b_{\tau \lambda} \partial_{\sigma}g_{\nu \xi} b_{\rho \chi} g^{\tau \xi} g^{\lambda \chi} - \frac{i}{4} \theta^{\mu \sigma} \partial_{\mu}b_{\tau \lambda} \partial_{\sigma}b_{\xi \chi} b_{\nu \delta} b_{\rho \alpha} g^{\tau \xi} g^{\lambda \delta} g^{\chi \alpha} \nn - \frac{i}{4} \theta^{\mu \sigma} \partial_{\mu} b_{\tau \rho} \partial_{\sigma}g^{\tau \xi} b_{\nu \xi} \nn \\ && + \frac{i}{4} \theta^{\mu \sigma} \partial_{\mu} b_{\nu \tau} \partial_{\sigma}b_{\lambda \rho}g^{\tau \lambda} + \frac{i}{4} \theta^{\mu \sigma} \partial_{\mu} g_{\nu \tau} \partial_{\sigma}g^{\tau \lambda} g_{\rho \lambda} + \frac{i}{4} \theta^{\mu \sigma} \partial_{\mu} b_{\tau \lambda} \partial_{\sigma}g^{\tau \xi} b_{\nu \chi} g_{\rho \xi} g^{\lambda \chi} + \mathcal{O}(\theta^2) \, . \nn \\
\eea
\end{widetext}
\noindent So far we have presented the construction of the vacuum Lagrangian in terms of the fundamental degrees of freedom, namely, the generalized dilaton and the generalized star metric. In the next part we will briefly discuss the inclusion of matter using a generalized scalar field.

\subsection{Inclusion of matter}
Now we focus on an $O(D,D)$ invariant free scalar field $\Phi$ coupled to the background content of a non-commutative DFT. For simplicity we consider the massless case, the parametrization of which is given by the ordinary scalar field $\phi$. The matter Lagrangian is given by,
\bea
{\cal L}_{\textrm{matter}}[ {\cal H}, \Phi] = \frac12   \partial_{ M}  \Phi \star {\cal H}^{\star M  N} \star \partial_{ N}  \Phi \, - V(\Phi) .
\label{Scalarlm}
\eea
To first order in $\theta$, the generalized star metric ${\cal H}^{\star \mu \nu}$ contains a $b$-field contribution. However, it is easy to see that the term depending on the $b$-field vanishes because of the symmetric contraction given by the derivatives of the scalar field. This effect happens order by order and we recover exactly the expression (\ref{scalargr}). The inclusion of scalar field dynamics in double non-commutative geometry is very promising because of the correspondence between these dynamics and the dynamics of a perfect fluid (see for example \cite{EyN1}-\cite{EyN3}), the latter also studied in non-commutative scenarios \cite{NCfluid1}-\cite{NCfluid3}. Imposing the generalized version of the correspondence given by
\bea
 U_{ M} & = & \frac{\partial_{M} \Phi}{\sqrt{|\partial_{P}\Phi \star {\cal H}^{\star PQ} \star \partial_{Q} \Phi|}} \, , \\
\tilde p & = & - \frac12 \partial_{P}\Phi \star {\cal H}^{\star PQ} \star \partial_{Q} \Phi - V(\Phi) \, , \\ 
\tilde e + \tilde p & = & |\partial_{P}\Phi \star {\cal H}^{\star PQ} \star \partial_{Q} \Phi| \, ,
\eea
in the energy momentum tensor of the generalized scalar field
\bea
{\cal T}_{MN} = && \eta_{MN} {\cal L}_m +4\, \overline{P}^{\star}_{[MK} \star P^{\star}_{N]L} \left(\frac{\delta {\cal L}_m}{\delta {P}^{\star}_{KL}}-\frac{\delta {\cal L}_m}{\delta \overline{P}^{\star}_{KL}}\right)\label{TMNlagrangian} \, , \nn
\eea
we find that the generalized energy momentum tensor for a perfect fluid coupled to the non-commutative double geometry is given by
\bea
{\cal T}_{ M  N} & = & 4\, \overline{P}^{\star}_{[M|K} \star P^{\star}_{N]L}(\sqrt{\tilde e + \tilde p} \ U_{M} \star \sqrt{\tilde e + \tilde p} \ U_{N}) \nn \\ && - \frac{1}{2} \eta_{M N} \sqrt{\tilde e + \tilde p} \ U_{P} \star {\cal H}^{\star P Q} \star \sqrt{\tilde e + \tilde p} \ U_{Q} \,
\label{em}
\eea

The previous generalized energy-momentum tensor describes statistical matter (perfect fluid dynamics) in the double geometry, and contains a tower of higher-derivative terms related to the $\theta$ contributions. This object can be used in the RHS of the generalized Einstein equation, which in turn corresponds with the EOM of $\cal H^{\star}_{M N}$. We leave this issue for future work.   

\section{Discussion and outlook}
\label{sectionDANDO}

In this work we have extended the DFT construction in order to include non-commutative effects through a Moyal-Weyl product. We have deformed generalized diffeomorphism transformations in a consistent way: expanding explicitly the $\theta$-contributions gives the transformation rules of the fundamental fields. We expect $\theta \propto \hbar$, and therefore the symmetries are preserved at the classical level. We catalog here the main differences between our construction and the standard (or classical) construction of DFT:
\begin{itemize}
\item The fundamental fields of the theory are given by the generalized star metric and the generalized dilaton. While both fields might have their own $\theta$-expansion, the former has to include contributions in order to guarantee Eq. (\ref{Oddnc}).

\item The projectors $P^{\star}$ and $\overline P^{\star}$ are no longer symmetric due to the $\theta$-expansion of the generalized star metric. This is similar to the antisymmetric deformation of the inverse metric in the non-commutative GR context. 

\item Both the duality transformations and the generalized diffeomorphisms are deformed, and as such they introduce non-covariant terms in the action from the point of view of ordinary DFT. Since the $b$-field is part of the components of the generalized star-metric, the Abelian gauge transformations are not preserved when $\theta\neq 0$.

\item The construction of (\ref{Hmetric}) can be done considering the generalized frame formalism of DFT \cite{FrameDFT1}-\cite{FrameDFT2}. In that case, the form of (\ref{Hmetric}) remains the same, but $g^{\mu \nu} = e^{\mu}{}_{a} \star e^{\nu a} + e^{\nu}{}_{a} \star e^{\mu a}$ and $g_{\mu \nu} = e_{\mu a} \star e_{\nu}{}{}^{a}  + e_{\nu a} \star e_{\mu}{}^{a}$. This means that ${\cal H}_{M N}$ might contain its own $\theta$-expansion.

\end{itemize}

We finish this section discussing some future directions:

\begin{itemize}

\item Covariant deformations: While the present analysis is based on a deformation of the generalized symmetries from the point of view of the commutative DFT, one can try to preserve the covariance of the star product using the covariant derivative as \cite{Harikumar}
\begin{equation}
F\star G\,(X) = e^{\frac{i}{2}\theta \nabla_{X} \otimes
\nabla_{Y}} F(X) \otimes G(Y)
\Big|_{Y\rightarrow X} \, .
\end{equation}
The main difficulty of these kinds of deformations is that on the one hand the resulting product is not associative anymore and, on the other hand, one has the problem of the undetermined connection $\Gamma_{M N P}$. However it would be interesting to study if these undetermined projections vanish from the resulting action.

\item $L_{\infty}$ structure and Hopf algebras: The $L_{\infty}$ structure of DFT is well known \cite{OZ} in the commutative case. Extending these studies to include the non-commutative case might be adequate to find straightforward generalizations to braided symmetries \cite{Braided} from the $L_{\infty}$ algebra \cite{Toni}-\cite{Blumen3}. Furthermore from the rewriting of the diffeomorphisms in the double geometry given in (\ref{rewriting}) it is possible to study generalizations of the standard Hopf algebra from a T-duality invariant perspective \cite{book}.

\item $\alpha'$-corrections: The terms produced by the $\circ$- and $\star$-product expansion are higher-derivative corrections. In case of covariant deformations, these terms could be related to the structure of higher- derivative corrections of DFT (see \cite{Electures} for a review) as in \cite{HSZ1}-\cite{HSZ3}. Exploring the interplay between non-commutativity and covariant higher-corrections might be an alternative way of avoiding some present obstructions of the double geometry as noticed in \cite{Linus}.

\item Non-Riemannian geometries: One possibility is to extend our formalism in order to describe non Riemannian geometries such as stringy Newton-Cartan geometry \cite{NC1}-\cite{NC2}. This means that the generalized metric now is parametrized considering an $(n,\bar{n})$ decomposition as in \cite{ParkNR1}-\cite{ParkNR2}. Since T-duality along the longitudinal direction of this theory describes a relativistic string theory on a Lorentzian geometry with a compact light-like isometry \cite{NRTD1}-\cite{NRTD2}, it is to be expected to find within a DFT a universal encoding of both Riemannian and non-Riemannian geometries. 
\end{itemize}

\subsection*{Acknowledgements}
We are indebted to L. Jonke and M. Dimitrijević Ćirić for enlightening discussions and detailed observations in the first version of the draft. We are also very grateful to Athanasios Chatzistavrakidis, F. Hassler and D. Marques for interesting comments and suggestions. The work of T. Kodžoman is supported by the Croatian Science Foundation project IP-2019-04-4168.

\end{document}